\documentclass[12pt,epsbox]{article}
\usepackage[dvips]{graphicx}

\makeatletter
\@addtoreset{equation}{section}
\renewcommand{\theequation}{\thesection.\arabic{equation}}

\newcommand{\nn}{\nonumber \\}

\newcommand{\tr}{{\rm tr}\,}

\newcommand{\be}{\begin{equation}} \newcommand{\ee}{\end{equation}}
\newcommand{\bea}{\begin{eqnarray}} \newcommand{\eea}{\end{eqnarray}}

\newcommand{\lef}{l_{\rm eff}}
\newcommand{\aef}{a_{\rm eff}}

\font\zfont = cmss10 
\newcommand{\ZZ}{\hbox{\zfont Z\kern-.4emZ}}

\ifx\TwoupWrites\UnDeFiNeD\else\target{\magstepminus1}{11.3in}{8.27in}
\source{\magstep0}{7.5in}{11.69in}\fi
\newfont{\fourteencp}{cmcsc10 scaled\magstep2}
\newfont{\titlefont}{cmbx10 scaled\magstep3}
\newfont{\authorfont}{cmcsc10 scaled\magstep1}
\newfont{\fourteenmib}{cmmib10 scaled\magstep2}
\skewchar\fourteenmib='177
\newfont{\elevenmib}{cmmib10 scaled\magstephalf}
\skewchar\elevenmib='177
\makeatletter
\newcommand\nonsequentialeqnum{
\@addtoreset{equation}{section}
\def\theequation{\arabic{section}.\arabic{equation}}}
\newif\ifp@bblock \p@bblocktrue
\newcommand\nopubblock{\p@bblockfalse}
\newcommand\topspace{\hrule height 0pt depth 0pt \vskip}
\newcommand\p@bblock{\begingroup \tabskip=\hsize minus \hsize
\baselineskip=1.5\ht\strutbox \topspace-2\baselineskip
\halign to\hsize{\strut ##\hfil\tabskip=0pt\crcr
\the\Pubnum\crcr\the\date\crcr}\endgroup}

\renewcommand\titlepage{\ifx\TwoupWrites\UnDeFiNeD\null
\vspace{-1.7cm}\fi
\vskip0.6cm
\ifp@bblock\p@bblock \else\hrule height 0pt \relax \fi}
\makeatother
\newtoks\date
\newtoks\Pubnum
\newtoks\pubnum
\newcommand{\frontpageskip}{\vspace{12pt plus .5fil minus 2pt}}
\renewcommand{\title}[1]{\frontpageskip
\begin{center}{\titlefont #1}\end{center}\par}
\renewcommand{\author}[1]{\frontpageskip\par\begin{center}
{\authorfont #1}\end{center}
\nobreak
}

\renewcommand{\thanks}[1]{\footnote{#1}}
\renewcommand{\abstract}{\par\frontpageskip\centerline{
\fourteencp Abstract}
\vspace{8pt plus 3pt minus 3pt}}

\begin{document}

\begin{titlepage}
\hfill
\vbox{
    \halign{#\hfil         \cr
           CERN-TH/2002-160 \cr
           TAUP-2707-02 \cr
           hep-th/0207223  \cr
           } 
      }  
\vspace*{20mm}
\begin{center}
{\Large {\bf  Penrose Limit and Six-Dimensional Gauge Theories}\\} 
\vspace*{15mm}
{\sc Yaron Oz}$^{a\,b}$ 
\footnote{e-mail: {\tt yaronoz@post.tau.ac.il, Yaron.Oz@cern.ch}}
and {\sc Tadakatsu Sakai}$^{a}$
\footnote{e-mail: {\tt tsakai@post.tau.ac.il}}

\vspace*{1cm} 
{\it {$^{a}$ Raymond and Beverly Sackler Faculty of Exact Sciences\\
School of Physics and Astronomy\\
Tel-Aviv University , Ramat-Aviv 69978, Israel}}\\ 

\vspace*{5mm}
{\it {$^{b}$Theory Division, CERN \\
CH-1211 Geneva  23, Switzerland}}\\

\end{center}

\begin{abstract}

We study the Penrose limit of
the $(p,q)$ fivebranes supergravity background.
We consider the different phases of the worldvolume field theory
and their weakly coupled descriptions.
In the Penrose limit we get 
a solvable string theory and compute the spectrum. It
corresponds to states of the six-dimensional worldvolume
theory with large energy and 
large $U(1)$ charge.
We comment on the RG behavior of the gauge
theory.

\end{abstract}
\vskip 2cm

July 2002

\end{titlepage}

\newpage

\section{Introduction}

The pp-wave background 
\cite{pp} is one of the solvable string theory backgrounds
\cite{Metsaev}.  It has received much attention recently since 
it can
be obtained by taking a Penrose limit of AdS$_5\times S^5$ \cite{bmn}.
Via the AdS/CFT correspondence (for a review, see \cite{ads}),
it corresponds to a particular limit of  ${\cal N}=4$ SYM theory, where 
the number of colors $N$ is taken to infinity, the Yang-Mills coupling $g_{YM}$ is
kept fixed and one considers operators with infinite R-charge $J$ such
that $J/\sqrt{N}$ is fixed.
Since string theory is solvable one obtains the 
exact spectrum of anomalous dimensions of a particular set of operators
of the gauge theory.
This discussion has been generalized to a large number of other AdS/CFT
examples.

It is natural to ask whether the Penrose limit is a useful tool to analyse
particular sectors of non-conformal field theories
\cite{Corrado:2002wi,cobi2,brecher}.
Non-local theories in the Penrose limit of the dual string background
have been studied in \cite{pp;ls}.
In particular, the Penrose limit of the NS5-brane background
is the 
Nappi-Witten background \cite{Nappi} (see also \cite{Gomis,Kiri}), 
which is solvable.
The string spectrum and its interpretation via states of
little string theory has been discussed in  \cite{pp;ls}.

In this letter we will consider the $(p,q)$
fivebranes in the Penrose limit. 
The worldvolume theory at low-energy is a six-dimensional 
${\cal N}=(1,1)$ SYM with 
gauge group $SU(s)$, where $s$ is the greatest common divisor
of $p$ and $q$. The different $(p,q)$ theories are
characterized by different $\theta$  angles.
They were considered in
\cite{Witten;6d} (see also \cite{Kol}).
A dual string description has been studied in \cite{oz;new6}.
We will show that in the Penrose limit the string theory is solvable.

The letter is organized as follows. In section 2 we present
the $(p,q)$ fivebranes 
supergravity background in the decoupling limit
and 
discuss the phase structure.
In section 3 we take the Penrose limit
and compute the string spectrum.
It provided information on the high-energy spectrum of the field theory.
Section 4 is devoted to the consideration of more general null geodesics and
some discussion.

\section{Field theory and the supergravity background}

\subsection{Field theory}
We will consider the $(1,1)$ supersymmetric six-dimensional gauge
theories on the worldvolume of $(p,q)$ fivebranes in Type IIB string theory.
At low energy the theories reduce to super Yang-Mills (SYM) theories
with gauge group $SU(s)$ where $s$ is the greatest common divisor of $p$
and $q$.  
Consider the low-energy effective action 
\bea
S_{eff} = \int {\rm d}^6x\left( \frac{1}{g_{YM}^2}\tr F^2 + \theta_{YM} \tr F \wedge F \wedge F +
\cdots\right) \ ,
\label{action}
\eea
where the dots correspond to higher dimension operators as well as the supersymmetric
completion.
Clearly, the low energy interactions dominated by the $F^2$ term cannot distinguish
the different  $(p,q)$ theories with the same low-energy gauge group.
However, it was argued in \cite{Witten;6d} that
the theta term in (\ref{action}), which is  a consequence
of $\pi_5(SU(s)) = {\bf Z}, s > 2$, is an observable
that distinguishes the low-energy gauge theories.
Our aim is to gain some information on the spectrum of these theories.

\subsection{Supergravity and phase structure}

Consider the supergravity (string) background 
\begin{eqnarray}
l_s^{-2}ds^2&\!=\!&{\hat h}^{-1/2}\left[-d x_0^2+\sum_{i=1}^{5}
d x_i^2+{ q \over u^2}(du^2+u^2d\Omega^2_3)\right]\ ,\cr
&&\cr
e^{\phi}&\!=\!& g \frac{{\hat h}^{-1}}{a_{\rm eff} u}\ , \quad
\chi={1\over g} a_{\rm eff}^2u^2 {\hat h}+\frac{p}{q}
\ , \cr
&&\cr
l_s^{-2}dB &\!=\!& 2q\,\epsilon_3 \ ,\quad
l_s^{-2}dA = 2p\,\epsilon_3 \ ,
\label{mypq}
\end{eqnarray}
where ${\hat h}^{-1}=1+a_{\rm eff}^2u^2$, $\aef^2=2\lef^2/q$
and $g l_{eff}^2 = g_s l_s^2$.
$\epsilon_3$ is the volume 3-form of $S^3$.
This background is obtained by taking a decoupling limit of the $(p,q)$
fivebranes supergravity solution
\cite{oz;new6} \footnote{Here we correct the form of the RR 2-form potential
$A$ 
of \cite{oz;new6}.}
\footnote{The IIB SUGRA solution of $(p,q)$ fivebranes is obtained
also in \cite{roy}.}.

String theory is weakly coupled if the effective string coupling 
is small $e^\phi < 1$.
This implies that $g \ll 1$.
We have a weakly coupled string description in
the regime 
$g <\aef u < g^{-1}$.

Consider the low-energy regime $a_{\rm eff}u < g$. 
The effective string coupling is large and we should obtain a weakly coupled
S-dual description. Assume 
$g\ll\frac{p}{q}$ which means
$\chi={p\over q}$.
Using the S-duality transformation
\begin{equation}
\tau\rightarrow {a\tau+b\over c\tau+d},\quad
dB\rightarrow d\,dB+c\,dA,\quad
dA\rightarrow b\,dB+a\,dA \ , \quad
\label{S}
\end{equation}
with $ad-bc=1$ and $cp+dq=0$, we get the D5-branes background 
\begin{eqnarray}
ds^2 &=& l_s^{2}\left(\frac{u}{(g_{YM}^2 s)^{1/2}}
(-d x^{2}_0+\sum_{i=1}^{5}
d x_i^{2})+ \frac{(g_{YM}^2 s)^{1/2}}{u}
(du^2+u^2d\Omega^2_3)\right) \ , \nonumber\\
e^{\phi} &=& \frac{(g_{YM}^2 s u^2)^{1/2}}{s}, \quad
\chi={a\over c}, \nn
dB &=&0,\quad
dA =2sl_s^{2}\epsilon_3 \ .
\label{d5}
\end{eqnarray}
$\left(g_{YM}^2s\right)^{1/2}= {q^2\aef\over s g}$
and we rescaled $x$ by 
$\sqrt{q\over g_{\rm YM}^2s}$.
$s$ is the greatest common divisor of $p$ and $q$.
The S-dual string description is weakly coupled when
$p\sim q\sim s$.

In the regime 
$(g_{\rm YM}^2s)^{-1/2}<u<s(g_{\rm YM}^2s)^{-1/2}$,
the scalar curvature of (\ref{d5}) is small and the supergravity description 
is valid.
When  $u<(g_{\rm YM}^2s)^{-1/2}$, the dimensionless effective gauge coupling
$g_{\rm YM}^2 su^2$ is small and the perturbative SYM theory
description is valid.

In the high-energy regime $\aef u> g^{-1}$,
the effective string coupling is large and 
we need to consider the  S-dual description.
Since
$\chi={1\over g}$, 
$g$
has to  be rational in order to have a weakly coupled S-dual description.
We rescale the coordinate
\begin{equation}
x_0 \rightarrow \sqrt{q}\,x_0,~~x_i \rightarrow \sqrt{q}\,x_i,~~
ga_{\rm eff}u=e^{U}.
\label{transf;uv}
\end{equation}
Using the S-duality transformation (\ref{S})
with
$c+gd=0$
we get
\begin{eqnarray}
&&ds^2=R^{2}\left(
-dx_0^2+dx_i^2+dU^{2}+d\Omega_3^2\right), \nn
&&\cr
&&e^{\phi}=c^{2}e^{-U},
\quad \chi={a\over c},\nn
&&\cr
&&dB=2R^{2}\epsilon_3,
\quad
dA=\chi\,dB \ ,
\label{uv}
\end{eqnarray}
where
$R^{2}=dql_s^2$ and $l_s^2$ is rescaled by  
${g_s\over |c|}$.
This is an NS5-branes metric with a constant RR scalar and RR
2-form potential.

To summarize, there are three different backgrounds which are relevant.
At low energies it is the D5-branes background (\ref{d5}), at intermediate
energies it is the $(p,q)$ fivebranes background (\ref{mypq})
and at high energies
it is the NS5-branes background (\ref{uv}).

\section{The Penrose limit}

In the following we will
take the Penrose limit of the three phases discussed in the
previous section.
The resulting backgrounds after an appropriate change of variables
take the form
\begin{eqnarray}
&&ds^2=-4dX^+dX^--y^2(dX^+)^2
     +d\vec{r}^2+d\omega^2+dy^2+y^2d\varphi^2, \nn
&&\chi={\rm const},\quad\quad \phi={\rm const},\nn
&&dB=a\,dX^+\wedge dy_1\wedge dy_2,\quad\quad dA=b\,dX^+\wedge
dy_1\wedge dy_2 \ ,
\label{generic;pp}
\end{eqnarray}
and 
\begin{equation}
a=2\cos\gamma,\quad b=2\left(e^{-\phi}\sin\gamma+\chi\cos\gamma\right) \ .
\label{coef;ab}
\end{equation}
$a,b,\gamma$ are constants and $y_1 = y \sin\varphi, y_2 =y \cos\varphi$.
It is straightforward to check that (\ref{generic;pp}) with
(\ref{coef;ab}) is a solution to the Type IIB supergravity equations.

Consider the $(p,q)$ fivebranes background (\ref{mypq}).
We take the metric on $S^3$ to be
\begin{equation}
d\Omega_3^2=d\theta^2+\sin^2\!\theta\, d\phi_1^2
+\cos^2\!\theta\, d\phi_2^2 \ ,
\label{metric;s3}
\end{equation}
and rewrite the background in terms of the
new coordinate
\begin{equation}
x_0 \rightarrow \sqrt{q}\,x_0,~~x_i \rightarrow \sqrt{q}\,x_i,~~
ua_{\rm eff}=e^{U}.
\label{transf;pq}
\end{equation}
The background becomes
\begin{eqnarray}
ds^2&=&R^2{\hat h}^{-1/2}\left[-dx_0^2+\sum_{i=1}^{5}
dx_i^2+dU^{2}+d\Omega^2_3\right]\ ,\cr
&&\cr
\chi&=&{1\over g}\, e^{2U} {\hat h}+{p\over q}
\ , \;\;\;\;\;\;\;\;\;\;\;\;\;\;\;\;\;\;
e^{\phi}=g \frac{{\hat h}^{-1}}{e^{U}}\ , \cr
&&\cr
dB &=& 2R^2\;\epsilon_3 \ ,\quad\quad
dA ={p\over q}\,dB \ ,
\end{eqnarray}
where ${\hat h}^{-1}=1+e^{2U}$ and $R^2=ql_s^2$.

In order to take the Penrose limit 
we first consider a null geodesic.
As a null geodesic we can take
\footnote{
A more general class of null geodesics will be discussed in the
discussion section.}
$U=U_0={\rm const},~x_i=\theta=0,~x_0=\phi_2=\lambda$, with
$\lambda$ the affine parameter.
We define a coordinate transformation such that
it reduces to the null geodesic in the large $R$ limit;
\begin{eqnarray}
&&U=U_0+(1+e^{2U_0})^{-1/4}\,{\omega\over R},\quad
\theta=(1+e^{2U_0})^{-1/4}\,{y\over R}, \nn
&&x_i=(1+e^{2U_0})^{-1/4}\,{r_i\over R}, \nn
&&x_0=(1+e^{2U_0})^{-1/4}\,
      \left(x^++{x^-\over R^2}\right),\quad
\phi_2=(1+e^{2U_0})^{-1/4}\,
      \left(x^+-{x^-\over R^2}\right).
\end{eqnarray}
By taking the limit $q\rightarrow\infty$
or equivalently $R\rightarrow\infty$, we obtain
\begin{eqnarray}
ds^2&=&-4dX^+dX^--y^2(dX^+)^2
     +d\vec{r}^2+d\omega^2+dy^2+y^2d\phi_1^2, \nn
&&\cr
\chi&=&{1\over g}\, {e^{2U_0}\over 1+e^{2U_0}}
     +\frac{p}{q} \ , 
\;\;\;\;\;\;\;\;\;\;\;\;\;\;
e^{\phi}=g \frac{1+e^{2U_0}}{e^{U_0}}\ , 
\nn
&&\cr
dB &=& 2(1+e^{2U_0})^{-1/2}\,dX^+\wedge dy_1\wedge dy_2 \ ,
\quad\quad
dA =\frac{p}{q}\,dB \ ,
\label{penrose;pq}
\end{eqnarray}
where
\begin{equation}
x^{\pm}=(1+e^{2U_0})^{\pm 1/4}\,X^{\pm}.
\end{equation}



Consider next the Penrose limit of the D5-branes background (\ref{d5})
which provides  a good description at low energy.
Define the new coordinate
\begin{equation}
u\,{(g_{\rm YM}^2s)^{1/2}\over s}=e^{U},\quad
x_{0,\cdots,5}\rightarrow (g_{\rm YM}^2s)^{1/2}x_{0,\cdots,5}
\ .
\label{transf;d5}
\end{equation}
The background takes the form
\begin{eqnarray}
&&ds^2=R^{2}e^{U}\left(
-dx_0^{2}+\sum dx_i^{2}+dU^{2}+d\Omega_3^2
\right),~~
e^{\phi}=e^{U} \ , 
\end{eqnarray}
where $R^{2}=sl_s^{2}$.
The Penrose limit is defined by
\begin{eqnarray}
&&U=U_0+e^{-U_0/2}\,
{\omega\over R},\quad
\theta=e^{-U_0/2}\,{y\over R},\quad
x_i=e^{-U_0/2}\,{r_i\over R},\nn
&&x_0=e^{-U_0/2}\,\left(
x^{+}+{x^{-}\over R^{2}}\right),\quad
\phi_2=e^{-U_0/2}\,\left(
x^{+}-{x^{-}\over R^{2}}\right) \ ,
\end{eqnarray}
with $R\rightarrow\infty$.
Using $x^{\pm}=e^{\pm U_0/2}X^{\pm}$
we obtain
\begin{eqnarray}
&&ds^{2}=-4dX^{+}dX^{-}
-y^{2}\left(dX^{+}\right)^2
+d\vec{r}^{2}+d\omega^{2}+d\vec{y}^{2}, \nn
&&e^{\phi}=e^{U_0},\nn
&&dB=0,\quad\quad 
dA=2\,e^{-U_0}
dX^{+}\wedge dy_1\wedge dy_2 \ .
\label{penrose;d5}
\end{eqnarray}


Finally, consider the NS5-branes background
(\ref{uv}) which is valid in the UV regime.
The Penrose limit is defined by the coordinates transformation
\begin{eqnarray}
&&U=
U_0+{\omega\over R},\quad
\theta={y\over R}, \quad
x_i={r_i\over R}, \nn
&&\cr
&&x_0=X^{+}+{X^{-}\over R^{2}},
\quad
\phi_2=X^{+}-{X^- \over R^{2}} \ ,
\end{eqnarray}
with $R\rightarrow\infty$.
One gets
\begin{eqnarray}
&&ds^{2}=-4dX^{+}dX^{-}
-y^{2}\left(dX^{+}\right)^2
+d\vec{r}^{2}+d\omega^{2}+d\vec{y}^{2}, \nn
&&e^{\phi}=c^{2}e^{-U_0},\quad\quad
\chi={a \over c}, \nn
&&dB=2\,dX^{+}\wedge dy_1\wedge dy_2,\quad\quad 
dA =2\chi dX^{+}\wedge dy_1\wedge dy_2 \ .
\label{penrose;uv}
\end{eqnarray}

\section{String spectrum}

In this section, we quantize string theory 
on the pp-wave backgrounds obtained in the previous section. 
For simplicity we focus on bosonic states.
The bosonic part of the Green-Schwarz action takes the form
\begin{equation}
2\pi\alpha^{\prime}S_b=
\int d\tau\int_0^{2\pi\alpha^{\prime}p^+}d\sigma\left(
\,{1\over 2}G_{MN}\,\partial_aX^M\partial^aX^N
+B_{MN}\partial_{\tau}X^M\partial_{\sigma}X^N
\right).
\end{equation}

Let us first consider the background (\ref{penrose;pq}).
$B$ is given by
\begin{equation}
B=2\,(1+e^{2U_0})^{-1/2}\mu y_2dX^+\wedge dy_1 \ ,
\end{equation}
where we rescaled $X^{\pm}\rightarrow \mu^{\pm 1} X^{\pm}$.
In the light-cone gauge $X^+=\tau$ the action reads 
\begin{equation}
2\pi\alpha^{\prime}{\cal L}_b={1\over 2}\left[
(\partial_a\vec{r})^2+(\partial_a\omega)^2+(\partial_a \vec{y})^2
-\mu^2\vec{y}^2\right]
+2\,(1+e^{2U_0})^{-1/2}\mu\,y_2\partial_{\sigma}y_1 \ .
\end{equation}
$\vec{y}$ and $\omega$ are massless free bosons.
The equations of motion for $y_1,y_2$ are given by
\begin{eqnarray}
&&\partial_a^2y_1+\mu^2\,y_1
+2\,(1+e^{2U_0})^{-1/2}\mu\,\partial_{\sigma}y_2=0, \nn
&&\partial_a^2y_2+\mu^2\,y_2
-2\,(1+e^{2U_0})^{-1/2}\mu\,\partial_{\sigma}y_1=0 \ .
\end{eqnarray}
We solve these equations by 
Fourier expanding $y_1,y_2$
\begin{equation}
y_i(\sigma,\tau)=\Sigma_n\alpha_n^i(\tau)e^{in\sigma/\alpha^{\prime} p^+} \ .
\end{equation}
The coefficients read 
\begin{eqnarray}
&&\alpha_n^1(\tau)=e^{-iw_n\tau}\tilde{\alpha}^+_n
+e^{+iw_{-n}\tau}\left(\tilde{\alpha}^+_{-n}\right)^{\ast}
+e^{-iw_{-n}\tau}\tilde{\alpha}^-_n
+e^{+iw_{+n}\tau}\left(\tilde{\alpha}^-_{-n}\right)^{\ast}, \nn
&&\alpha_n^2(\tau)=i\left[-e^{-iw_n\tau}\tilde{\alpha}^+_n
+e^{+iw_{-n}\tau}\left(\tilde{\alpha}^+_{-n}\right)^{\ast}
+e^{-iw_{-n}\tau}\tilde{\alpha}^-_n
-e^{+iw_{+n}\tau}\left(\tilde{\alpha}^-_{-n}\right)^{\ast}\right] \ ,
\end{eqnarray}
where
\begin{equation}
\omega_n=\sqrt{\mu^2
+{n^2\over (\alpha{\prime}p^+)^2}
+2\,(1+e^{2U_0})^{-1/2}\mu\,{n\over \alpha^{\prime}p^+}} \ .
\end{equation}
Upon quantization, $\tilde{\alpha}_n^{\pm \dagger}$ become operators
that create states of light-cone energy $\omega_{\pm n}$.
Defining the oscillators
\begin{equation}
\tilde{\alpha}_n^{\pm}={1\over\sqrt{4p^+\omega_{\pm n}}}\,\alpha_n^{\pm} \ ,
\end{equation}
we see that they obey the commutation relations
\begin{equation}
\left[\alpha_n^+,\,(\alpha_m^+)^{\dagger}\right]=
\left[\alpha_n^-,\,(\alpha_m^-)^{\dagger}\right]=i\delta_{n,m} \ .
\end{equation}
Thus, the bosonic part of the light-cone hamiltonian takes
the form
\begin{equation}
2p^-=\sum_n\left(N_n^{r,w}{|n|\over\alpha^{\prime}p^+}
+N_n^+\omega_n+N_n^-\omega_{-n}\right) .
\end{equation}

We would like 
to relate the string spectrum to states in the six-dimensional
field  theory.
Note that
\begin{equation}
{\partial\over\partial X^+}
={\partial\over\partial x_0}+{\partial\over\partial\phi_2},\quad
{\partial\over\partial X^-}={(1+e^{2U_0})^{-1/2}\over R^2}
\left({\partial\over\partial x_0}-{\partial\over\partial\phi_2}\right).
\end{equation}
It follows from this that
\begin{equation}
{2p^-\over\mu}=q^{1/2}E-J_V,\quad
2\mu p^+={(1+e^{2U_0})^{-1/2}\over R^2}(q^{1/2}E+J_V) \ ,
\end{equation}
where we used
\begin{equation}
q^{1/2}E=i{\partial\over\partial x_0},\quad 
J_V=-i{\partial\over\partial \phi_2} \ .
\end{equation}
Here we define the energy $E$ in terms of the generator of the
translation of $x_0$ in (\ref{mypq}).
We thus find that the string spectrum with finite $p^{\mp}$
corresponds to states in the six-dimensional field theory with
\begin{equation}
q^{1/2}E,J_V\sim q,~~{\rm while}~ E-J_V={\rm finite} \ ,
\end{equation}
with $q \rightarrow \infty$.
Note that for large $q$ 
\begin{equation}
R^2\mu p^+=(1+e^{2U_0})^{-1/2} J_V \ .
\end{equation}
Thus, the light-cone energy takes the
form
\begin{equation}
{2p^-\over\mu}=\sum_n\left[
(1+e^{2U_0})^{1/2}\,{q\over J_V}N_n^{r,w}|n|+
\sqrt{1+(1+e^{2U_0})\,{q^2\over J_V^2}\,n^2 +2\,{q\over J_V}n}
\,\,(N_n^++N_{-n}^-)
\right].
\label{energy;pq}
\end{equation}

Consider now symmetry of the pp-wave background (\ref{penrose;pq}).
The $(p,q)$ fivebranes background before the Penrose limit
has an $SU(2)_L\times SU(2)_R$ isometry
of $S^3$ corresponding to the $R$-symmetry of the field theory. 
In the parametrization (\ref{metric;s3}), the 
manifest $U(1)$ isometries are
\begin{equation}
U(1)_{1,2}:~\phi_{1,2}\rightarrow\phi_{1,2}+{\rm const} \ .
\end{equation}
They are identified as 
\begin{equation}
U(1)_1=U(1)_A,\quad U(1)_2=U(1)_V \ ,
\end{equation}
where $U(1)_{V(A)}$ is the vector (axial) subgroup of
$U(1)_L\times U(1)_R$.
Denote
\begin{equation}
Y=(y_1+iy_2)/2
=\sum_n\left[
{1\over\sqrt{4p^+\omega_n}}\,
\alpha_n^+e^{-i\omega_n\tau+in\sigma/\alpha^{\prime} p^+}
+{1\over\sqrt{4p^+\omega_{-n}}}\left(\alpha_n^-\right)^{\dagger}
e^{i\omega_{-n}\tau-in\sigma/\alpha^{\prime} p^+}
\right] \ .
\end{equation}
This complex scalar carries a $U(1)_A$ charge while the massless
scalars are neutral.

Recall that the string spectrum consists of states of the form
$( \omega, r)(Y,\bar{Y})|0\rangle$.
Thus, 
the corresponding states of the six-dimensional field theory
have energies  
varying differently with $n$ depending on whether
they carry $U(1)_A$ or not.

The computation of the string spectrum for the background 
(\ref{penrose;d5}) is straightforward:
since $dB=0$, the bosonic part of the GS action in the light-cone gauge
takes the
form
\begin{equation}
2\pi\alpha^{\prime}{\cal L}_b={1\over 2}\left[
(\partial_a\vec{r}^{\prime})^2+(\partial_a\omega^{\prime})^2
+(\partial_a \vec{y}^{\prime})^2
-\mu^{2}\vec{y}^{\prime 2}\right].
\end{equation}
The light-cone energy is
\begin{equation}
2p^{-}=
\sum_n\left(N_n^{r,w}{|n|\over\alpha^{\prime}p^{+}}
+N_n^{y}\sqrt{\mu^{2}
+{n^2\over (\alpha^{\prime}p^{+})^2}}
\right) .
\end{equation}
Note that
\begin{equation}
{2p^{-}\over \mu}=s^{1/2}E-J_V,\quad
2p^{+}\mu={e^{-U_0}\over R^{2}}\,(s^{1/2}E+J_V) \ ,
\end{equation}
and 
\begin{equation}
R^{2}p^{+}\mu=e^{-U_0}\,J_V \ .
\end{equation}
The light-cone energy reads
\begin{equation}
{2p^{-}\over\mu}=\sum_n\left[
e^{U_0}\,{s\over J_V}N_n^{r,w}|n|+
\sqrt{1+e^{2U_0}\,{s^2\over J_V^2}\,n^2}
\,\,N_n^{y}
\right] \ .
\label{energy;d5}
\end{equation}

We see that 
string states that carry non-vanishing $U(1)_A$ charge have
a different light-cone energy compared
with neutral states which are like 
states of strings in a flat background.

Finally consider the UV description (\ref{penrose;uv}).
The bosonic light-come energy is 
as that of little string theory (LST).
The spectrum of little string theory in the Penrose limit
was discussed in \cite{pp;ls}.
Note, however, that the background has now RR fields. Therefore, we expect
the interactions to differ from those of LST.

\section{Discussion}

In this letter, we discussed the Penrose limit of $(p,q)$
fivebranes and the the dual six-dimensional
field theory.
As in the case of \cite{bmn}, the limit yields a solvable
string background, and the string computation provides 
information on a subsector of states of the field theory with
large energy and large $U(1)$ charge.
It would be nice
to confirm the string computation from the viewpoint of
the six-dimensional field theory.
For this purpose, it could be useful to work with the matrix theory
formulation which was proposed in \cite{Witten;6d}.



We have focused on a particular class of null geodesics
such that the energy scale $U$ is constant.
Thus we are probing the field theory at a particular  energy scale.
One could be interested in a more general null geodesic 
with a non constant $U$. This is relevant to
the RG flow of the field theory.
Consider null geodesics of the $(p,q)$ fivebranes background
(\ref{mypq}).
$U$ must satisfy
\begin{equation}
\left( 1+e^{2U}\right)^{1/2} \dot{U}=\xi \ .
\end{equation}
Here $\cdot=d/d\lambda$ with $\lambda$ the affine parameter of the null
geodesic, and $\xi$ is a constant.
The solution reads
\begin{equation}
\sqrt{1+e^{2U}}-\tanh^{-1}\!\sqrt{1+e^{2U}}=
\xi\lambda+\eta \ ,
\end{equation}
with $\eta$ constant.
When $\xi=0$ it  reduces to the case
considered in the previous section.
The mass of the world-sheet scalar $\vec{r}$
in light-cone gauge is \cite{pp,cobi2}
\begin{equation}
m^2=-\tilde{h}^{1/4}\,{{d^2\tilde{h}^{-1/4}\over d\lambda^2}}
={\xi^2\over 4}\,{e^{2U}(e^{2U}-4) \over (1+e^{2U})^3} \ .
\end{equation}
The resulting string background is not 
solvable.
Note some interesting properties: first
$m^2$ is $x^+$-dependent. The light-cone
time-dependent mass in the context of holographic RG in Penrose
limits was noted by several authors 
\cite{hrg;pp,Corrado:2002wi,cobi2,brecher,pp;ls}.
Second, $m^2$ becomes negative in the IR regime $U\rightarrow 0$.
This occurs also in the
Penrose limits of Dp($p<5$) backgrounds  \cite{cobi2}.
However, in the IR the relevant background is 
(\ref{d5}).
The mass 
of $\vec{r}$ in the IR is 
\begin{equation}
m^2={\xi^{\prime 2}\over 4}\,
{1\over (\xi^{\prime}x^{+}+\eta^{\prime})^2} \ ,
\end{equation}
where $\xi^{\prime},\eta^{\prime}$ are constants.
Thus,  we have $m^2$ positive in the IR.
It would be interesting to study the Penrose limit for these
null geodesics in more detail 
and explore the RG behavior of the six-dimensional
field theory.

\section*{Acknowledgements}

We would like to thank J.~Sonnenschein for a valuable discussion.
Y. Oz would like to thank Queen Mary university and
the Newton Institute at Cambridge, and 
T. Sakai would like to thank the TH-division                                   at CERN for hospitality.
This research is supported by the US-Israel Binational Science
Foundation.

\newpage


\end{document}